\documentstyle[twocolumn,prl,aps]{revtex}
                    \begin{document}
                    \draft
                    \title{Tight Binding Hamiltonians and Quantum Turing
          Machines} 
                    \author{Paul Benioff\\
                     Physics Division, Argonne National Laboratory \\
                     Argonne, IL 60439 \\
                     e-mail: pbenioff@anl.gov}
                     \date{\today}

                    \maketitle
                    \begin{abstract}  
          This paper extends work done to date on quantum computation by
          association of potentials with different types of steps.  Quantum
          Turing machine Hamiltonians, generalized to include potentials,
          correspond to sums over tight binding Hamiltonians each with a
          different potential distribution. Which distribution applies is
          determined by the intitial state.  An example, which enumerates
          the integers in succession as binary strings, is analyzed. It is
          seen that for some initial states, the potential distributions
          have quasicrystalline properties and are similar to a
          substitution sequence. 
                    \end{abstract}

          \section{Introduction}

          Quantum computation is a field of much interest.  Since the early
          work \cite{Benioff1,Deu}, much impetus was provided by Shor's
          discovery \cite{Shor} that for some problems quantum computers
          are more efficient than classical computers.  Recent work
          includes discussion of quantum error correction \cite{Shor1}, use
          of quantum gates for computation \cite{Lloyd},  and Hamiltonian
          descriptions of quantum Turing machines \cite{Benioff2,Benioff}.  

          In all work to date models of quantum computation are considered
          in which successive steps of the computation proceed smoothly
          with no potentials present, if the system is isolated from the
          environment (see for example \cite{Lloyd1}).  Landauer
          \cite{Landauer} has noted that environmental influences and
          defects in physical models introduce random potentials which
          degrade performance by causing reflections at various steps and
          decay of the transmitted component.  

          Here potentials are introduced into quantum computation, not in a
          random fashion, but by association with different types of steps. 
          An example is the association of a potential with all Turing
          machine "read 1" steps. All other steps are potential free.  

          The method used is to generalize the description of quantum
          Turing machines \cite{Benioff1,Benioff} to include potentials at
          one or more types of computation steps. Isolation from the
          environment is assumed.  In this case the Feynman \cite{Feynman}
          Hamiltonian for these generalized quantum Turing machines
          (GQTM)s,  can be also be described as a sum of tight binding
          Hamiltonians each of which has a different potential distribution
          along a one dimensional path of computation states.  Which of
          these Hamiltonians applies to the GQTM is determined by the
          initial state. 

          For some GQTMs the potential distributions are similar to those
          which are of much recent interest in condensed matter physics. 
          This includes Hamiltonians with quasiperiodic potentials
          \cite{DiV1} or with potential distributions corresponding to
          substitution sequences (see \cite{defsub} for a definition).
          Examples include the Period Doubling, \cite{Luck,BoGh1}, Thue-
          Morse \cite{Luck,BoGh1}, Fibonacci \cite{Suto}, and Rudin-Shapiro
          \cite{BoGh2} substitution sequences.  This similarity provides a
          link between the behavior of a GQTM as it evolves during a
          computation and the 1-D motion of a particle such as an electron
          in systems with potentials which are quasiperiodic or correspond
          to substitution sequences.  

          This similarity will be shown here by first giving a brief
          description of the physical model and the sum decomposition of
          the GQTM Hamiltonian. An example, the counting GQTM which
          enumerates the nonnegative integers in succession as binary
          strings will be considered.  This is of interest because some of
          the potential distributions are similar to but not identical with 
 those for both
          quasicrystals and substitution sequences.  The differences
          suggest that the distributions are a new type of 1-D structure
          that does not seem to have been discussed in the literature.

          \section{The Physical Model}

          For quantum Turing machines, the physical model corresponds to
          one tape machines.  It includes a finite state head that moves
          along an infinite lattice of finite state systems called qubits.
          For many computations it is sufficient to consider binary strings
          corresponding to qubit states $\vert 0\rangle, \vert 1\rangle$. 
          The actual physical form of each qubit (e.g. as a spin 1/2
          system, etc.) is not important here.  

          The computation basis $B$ which spans a separable Hilbert space
          $\cal H$ is the set $\{\vert l,j,S\rangle \}$ of states.  Here
          $j$ and $l$ denote the lattice location and internal state of the
          head. $\vert S\rangle= \otimes_{j=-\infty}^{\infty}\vert
          S(j)\rangle$ denotes the lattice qubit state where  $S$ is any
          function from the integers into the set of possible qubit states
          such that $S(j) \neq 0$ for at most a finite number of $j$
          values. This condition, the $0$ tail state condition, is one of
          many that can be imposed to keep $B$ denumerable. Both $B$ and
          $\cal H$ depend on the condition imposed. 

          \section{Step Operators and Hamiltonians}

          Each GQTM is described by a step operator $T$ such that iteration
          of $T$ or its adjoint corresponds to successive steps of the GQTM
          in the forward or backward time direction.  $T$ is defined as a
          finite sum of elementary step operators, $T=\sum_{l,s}T_{l,s}
          =\sum_{l,s}\gamma_{l,s}W_{l,s}$, over head states $l$ and qubit
          states $s$ where $\gamma_{l,s}$ is a real constant between $0$
          and $1$.. $W_{l,s}$ has the form,
          \begin{equation}
          W_{l,s} = \sum_{j=-
          \infty}^{\infty}w_{l,s}Q_{l}v_{l,s}P_{s,j}u_{l,s}P_{j}
          \label{wls}
          \end{equation}
          where $Q_{l},P_{j},P_{s,j}$ are respective projection operators
          for the head in internal state $\vert l\rangle$, at site $j$, and
          the site $j$ qubit in state $\vert s\rangle$.  

          The unitary operators $w_{l,s},v_{l,s},u_{l,s}$ describe head and
          qubit state changes, and head motion.  The dependence on $l,s$ is
          indicated by the subscripts.  The operators $w_{l,s}$ and
          $u_{l,s}$ satisfy the commutation relations, $w_{l,s}Q_{l}
          =Q_{l^{\prime}} w_{l,s}$ for some $l^{\prime}\epsilon L$ and
          $u_{l,s}P_{j}=P_{j\pm 1}u_{l,s}$. Here $L$ is the finite set of
          head states and $P_{j+1},P_{j-1}$ correspond to head
          motion one site to the right or one site to the left.  For our
          purposes it is sufficient to require that $v_{l,s}$ satisfy
          $v_{l,s}P_{s,j}= P_{s^{\prime},j}v_{l,s}$ where
          $s^{\prime}\epsilon \cal S$, the set of qubit states. If the
          qubits are binary this is equivalent to requiring that
          $v_{l,s}=\sigma_{x}$ or $v_{l,s}=1$.

          These step operators generalize earlier definitions
          \cite{Benioff1,Benioff2,Benioff} in that values of $\gamma_{l,s}$
          different from $0$ or $1$ are allowed. Potentials are associated
          with steps for which terms $T_{l,s}$ with $\gamma_{l,s} <1$
          are active. 

          The  main condition imposed on $T$ is that it be distinct path
          generating on the computation basis $B$.  This condition can be
          relaxed to apply to any basis, not just $B$ (see \cite{Benioff}
          for details) but this generality is not needed here. The
          condition means that iteration of $T$ or $T^{\dag}$ on the states
          in $B$ generates finite or infinite paths of states which do not
          join, branch, or intersect.   Mathematically this is expressed by
          requiring that $T$ be a direct sum of weighted shifts
          \cite{Halmos}.  That is $T=UD=\sum_{i}U_{i}D_{i}$ where for each
          $i$ $U_{i}$ is a bilateral shift, a unilateral shift, the adjoint
          of a unilateral shift, a finite truncated shift of length $N$, or
          a cyclic shift of length $M$ with $N,M$ arbitrary.    $D$ is
          defined by $D\vert l,j,S\rangle = \gamma_{l,S(j)}\vert
          l,j,S\rangle$ for any state $\vert l,j,S\rangle$ in $B$ not
          annihilated by $U$ or $U^{\dag}$.   Both $U$ and $D$ satisfy
          $U=\sum_{i}U_{i}P_{i}, D=\sum_{i}D_{i}P_{i}$ where the $P_{i}$
          are orthogonal multidimensional projection operators on $\cal H$.

          The $i$ sum is a sum over paths where each path $i$ in $B$ is
          defined by iteration of $U_{i}$ or $U_{i}^{\dag}$.  That is if
          the state $\vert l,j,S\rangle$ is not annihilated by $U_{i}$ or
          $U_{i}^{\dag}$ then path $i$ is defined by $\cdots
          U_{i}^{\dag}\vert l,j,S\rangle ,\vert l,j,S\rangle ,U\vert
          l,j,S\rangle ,\cdots$.

          For each step operator $T$ define a Hamiltonian by
          \cite{Feynman},
          \begin{equation}
          H=K(2-T-T^{\dag}) \label{ham}
          \end{equation}
          where $K$ is an arbitrary constant.  If $T=UD$, then $H$ can also
          be written as $H=\sum_{i}H_{i}P_{i}$ where $H_{i}=K(2-T_{i}-
          T^{\dag}_{i})$ describes motion along the path states in $B_{i}$.
          Eq.\ref{ham}  is based on the association of an infinitesimal
          time interval with $T$ so that it can be directly used to
          construct a Hamiltonian.

          From this it follows that for each path $i$, $H_{i}$ is a tight
          binding Hamiltonian on $P_{i}\cal H$ with nearest neighbor off
          diagonal potentials.  To see this note that
          \begin{equation}
          H_{i}=K(2-U_{i}-U^{\dag}_{i}) + V_{i} \label{tbhi}
          \end{equation} 
          where the off-diagonal potential $V_{i}$ along path $i$ is given
          by $V_{i}=K[U_{i}(1-D_{i})+(1-D_{i})U^{\dag}_{i}]$. The constant
          term $2K$ is present so that for each path, the path kinetic
          energy component of $H_{i}$, $K(2-U_{i}-U^{\dag}_{i})$ is
          equivalent to the symmetrized discrete version of the second
          derivative, $-K d^{2}/(dx)^{2}$. 

          Under the action of $H$ the initial state $\Psi(0)$ determines
          which tight binding Hamiltonian applies.  If $\Psi (0)$ is a
          linear sum of computation states in $P_{i}\cal H$, then $H_{i}$
          gives the evolution of $\Psi(0)$.  If $\Psi(0)$ has components in
          more than one subspace $P_{i}\cal H$, then each component has a
          corresponding $H_{i}$ which describes its evolution.  

          For many $i$ the potential distributions in the corresponding
          $H_{i}$ are the same.  This is the case for any transformation
          group whose actions do not change $T$.  For example, for GQTMs,
          each shift of an initial state along the lattice corresponds to
          a different path $i$ with the same $H_{i}$.   From this one sees that
          equivalence classes ${\cal P}_{i}=\{P_{j}:H_{j}=H_{i}\}$ of
          projectors can be defined which  identify paths with the same
          $H_{i}$.

          It is important to distinguish between path motion and head
          motion on the qubit lattice.  In general, iterations of $T$
          describe head motion backward and forward over lattice sites.
          However, (cyclic paths excepted) for each passage over a site,
          the overall system state is orthogonal to the state for any other
          passage. As a result the back and forth lattice motion can be
          unfolded to motion along a path where the path states are
          pairwise orthogonal and are ordered by iteration of $T$ or
          $T^{\dag}$.  The tight binding Hamiltonians, Eq. \ref{tbhi}, 
          refer to motion along paths and not the qubit lattice.   

          \section{The Counting GQTM}

          Simple examples can be constructed by letting $\gamma_{l,s}$
          assume just two values, $1$ and $\gamma$ where $\gamma$ is a real
          number between $0$ and $1$.  In particular, for all $l\epsilon
          L$, let $\gamma_{l,1}=\gamma$ and  $\gamma_{l,s}=1$ for all $s
          \neq 1$. 

          This simplification corresponds to GQTMs in which a potential is
          associated with all "read 1" elementary steps (those steps with
          $P_{sj} =P_{1j}$ in Eq. \ref{wls}). All other qubit state read
          actions are potential free.  In this case the $l,s$ sum for $T$
          can be split into two parts, $T=\sum_{l,s \neq 1}W_{l,s} +\gamma
          \sum_{l} W_{l,1}$.

          The counting GQTM can be used to illustrate the foregoing.  This
          GQTM, starting with a blank qubit lattice (all $0s$) with two
          markers spaced n+1 sites apart, generates all binary numbers in
          succession from $0$ up to $2^{n}-1$ by continually adding $1$. 

          The step operator for this GQTM, which is used in Eq. \ref{ham},
          is a sum of 7 terms:
          \begin{eqnarray}
          T & = & \sum_{j=-\infty}^{\infty}(Q_{0}P_{0j}uP_{j}
          +wQ_{0}P_{2j}uP_{j}+Q_{1}P_{0j}uP_{j}   \nonumber \\
           & & \mbox{}+wQ_{1}P_{2j}u^{\dag}P_{j} +\gamma
          Q_{2}v_{xj}P_{1j}u^{\dag}P_{j}+w^{\dag}Q_{2}v_{xj}P_{0j}uP_{j}  
          \nonumber \\
           & & \mbox{}+wQ_{2}P_{2j}uP_{j}) \label{Tex}
          \end{eqnarray}
          The projection operators are as defined for Eq. \ref{wls}; $w$ is
          a shift mod 3 on the three head states ($wQ_{m}=Q_{m+1}w \bmod
          3$) and $u$ shifts the head along the lattice by one site
          ($uP_{j}=P_{j+1}u$).  The need for markers is accounted for here
          by choosing the qubits in the lattice to be ternary with states
          $\vert 0\rangle ,\vert 1\rangle ,\vert 2\rangle$. $\vert
          2\rangle$ is used as a marker and $\vert 0\rangle ,\vert
          1\rangle$ are used for binary strings.  The qubit transformation
          operator for the site $j$ qubit $v_{xj}=\sigma_{xj}(P_{0j}+P_{1j})+
          P_{2j}$ exchanges the
          states $\vert 0\rangle , \vert 1\rangle$ and leaves the
          state $\vert 2\rangle$ alone.  

          The terms in $T$ are chosen so that $T$ is distinct path generating 
          on $B$. The theorems given in \cite{Benioff}, that verify this 
conclusion were proved for $D=1$.  The proofs should also hold for $D\neq 1$.

          An initial state for this GQTM is shown in Figure 1 with the head
          in state $\vert 0\rangle$ in a wave packet localized to the left
          of the origin.  All qubits are in state $\vert 0\rangle$ except
          those at sites $0,n+1$ which are in state $\vert 2\rangle$. 
          Successive iterations of $T$ move the head to the righthand units
          marker followed by successive enumeration of the first $2^{n}$
          numbers as binary strings. When the space between the markers is
          filled with $1s$, the initial state is restored by conversion of 
          the $1s$ to $0s$.  The head
          in state $\vert 1\rangle$ moves to the right with no more qubit
          or head state changes. 

          For this GQTM, a potential of height $V=2K(1-\gamma )$ is present
          at path states at which term 5 is active. This is the only term
          with a "read 1" operator.  The path potential distribution is
          determined by a close examination of successive iterations of
          $T$.  For example, to add $1$ to $\cdots 01001112$ to obtain
          $\cdots 01010002$ with the head beginning and ending in state
          $\vert 1\rangle$ at the units marker $2$ requires one "read 2"
          step (term 4) followed by three "read 1" plus $1\rightarrow 0$ 
          steps (term 5) followed by one "read 0" plus $0\rightarrow 1$
          step (term 6) followed by three "read 0" steps (term 3). In this
          path segment there is a potential of height $V$ and width $3$
          with potential free regions of width $1$ to the left and
        width $4$ to the right.  Extension of the segment shows a few more
          potential free sites in both directions before encountering the
          next potential.

          The path potential distribution is expressed in general as
          follows:  Let $R$ be a function from the nonnegative integers to
          the nonnegative integers such that $R(j)$ gives the number of
          $1s$ occurring before the first $0$ in the binary string for $j$. 
          $R(j) =0$ if $j$ even and $R(1)=1,\: R(3)=2,\: R(5)=1,\:
          R(7)=3,\cdots$.  $R$ can be expressed recursively by 
          \begin{equation}
          R_{n}=S_{n-1}n;\;\;  S_{n}=R_{n}S_{n-1} \label{Rrecur}
          \end{equation}
          for $n=1,2,\cdots$ with $S_{0}=0$ where $R_{n}$ is the initial
          segment of $R$ of length $2^{n}$.

          It turns out that $R$ is a substitution sequence \cite{defsub}
          with the substitution rule $n\rightarrow 0n+1$ for
          $n=0,1,2,\cdots$. (That is, the infinite sequence 
	  $0,1,0,2,0,1,0,3,\cdots$
          is invariant under the replacement of $n$ with $0n+1$.)  This
          generalizes the literature definition of substitution sequences 
          \cite{BoGh1,KoNo,Queff} in that the alphabet is infinite.

          A study of the iteration of $T$ with the initial head state
          $\vert 1,n+1\rangle$ and $\vert S\rangle$ as shown in Fig. 1
          shows that the potential distribution is given by a binary
          sequence $\cal R$  obtained from  $R$ by replacing $R(j)$
          by a string of numbers according to the following prescription: 
          $0\rightarrow 00=\underline{0}$ and $n\rightarrow
          0n0^{n+1}=\underline{n}$. Here $0^{n+1}$ denotes a string of
          $n+1$ $0s$ and $n$ denotes a string of $n$ $1s$.  Justification
          for this can be seen by extension of the example of adding $1$ to
          $\cdots 01001112$. 

          $\cal R$ is generated recursively by Eqs. \ref{Rrecur} if $n$ is
          replaced by $\underline{n}$ and $S_{0}=\underline{0}$. It follows
          that $\cal R$ is also a substitution sequence for the words
          $\underline{0}, \underline{1},\cdots$ regarded as an alphabet.
          The recursive generation is important for future work on the
          spectral and transmission properties of tight binding
          Hamiltonians with this potential distribution.  

          The above is illustrated in Figure 2 for $n=6$.  The figure shows
          the distribution of read 0 and read 2 steps (as $0s$) and read 1
          steps (as $1s$) as a function of the number of $T$ iterations
          starting from the initial head state $\vert 1,n+1\rangle$ and the
          two markers separated by $6$ sites (Figure 1).  Fig. 2 also gives
          the potential distribution for enumerating the first 64 numbers
          as binary strings.  Changing the initial state to any component
          of the wave packet shown in Figure 1 shifts the distribution
          along the path with no changes in the width or spacing between the 
          potentials.

          Different initial states correspond to different potential
          distributions. Changing the separation $n+1$ between the two
          markers changes the number of potentials present. Initial states
          with $m$ markers separated by $n_{1}+1, n_{2}+1,\ldots ,n_{m}+1$ qubit
          sites  give a potential distribution obtained
           by concatenating (with some intervening $0s$)
          initial segments of $\cal R$ corresponding to enumeration of the 
first $2^{n_{1}},2^{n_{2}},
          \ldots ,2^{n_{m}}$ numbers. If all the $n_{i}=p$ are the same, the
          potential distribution is periodic with $m$ periods each
          containing the first $2^{p-1}$ potentials.  

          If just one marker is present and the head is in state $\vert
          1\rangle$, then all binary numbers are generated in succession
          without halting.  For this input state the potential distribution
          in $H_{i}$, Eq. \ref{tbhi}, is given by $\cal R$. The system is
          quasicrystalline in that no two path sites have the same global
          potential environment, yet each local environment of arbitrary
          length in  $\cal R$ is repeated infinitely often \cite{DiV}.
          Differences between this distribution and those described in the
          literature include the facts that the potential distribution is
          one-way infinite, the alphabet is infinite, and $\cal R$ is a
          substitution sequence with respect to alphabet words and not the
          original alphabet.  Because of this the 1-D structure appears to
          be of a new type not yet analyzed.

          \section*{Acknowledgements}
          This work is supported by the U.S. Department of Energy, Nuclear 
          Physics Division, under contract W-31-109-ENG-38.

          \begin{center}
          FIGURE CAPTIONS
          \end{center}

          Figure 1.  Initial and Final States for Counting GQTM for the
          First $2^{n}$ Binary Numbers.  All lattice qubits are in state
          $\vert 0\rangle$ except those at sites $0$ and $n+1$ which are in
          state $\vert 2\rangle$.  The initial and final head states are
          shown as wave packets with internal head states $\vert 0\rangle$
          and $\vert 1\rangle$ to the left and right respectively. \\
          \\
          Figure 2. Path Potential Distribution for Enumeration of the
          first 64 Numbers ($n=6$) as Binary Strings.  The figure shows the
          distribution of $0s$, corresponding to read 0 or read 2 steps,
          and $1s$, corresponding to read 1 steps, as a function of the
          number of $T$ iterations starting from the initial head state
          $\vert 1,n+1\rangle$.


\begin{thebibliography}{99}
\bibitem{defsub}
          Let $A$ be a finite alphabet, $A^{*}$ the set of all finite words
          of $A$, and $A^{N}$ the set of all one way infinite words of $A$.
          A substitution rule replaces each letter $\alpha$ of $A$ with a
          word from $A^{*}$. For each $\alpha$ a sequence of words in
          $A^{*}$ can be generated by starting with $\alpha$ and iterating
          the substitution rule.  The limit sequence in $A^{N}$ is a
          substitution sequence if it is invariant under the rule.
\end{thebibliography}
          
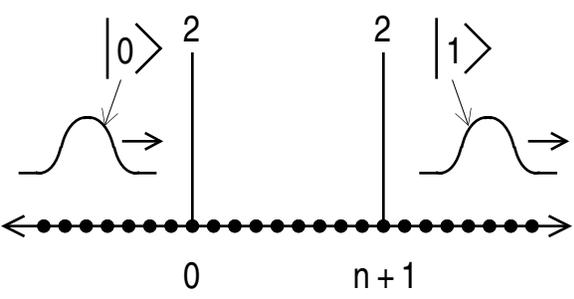
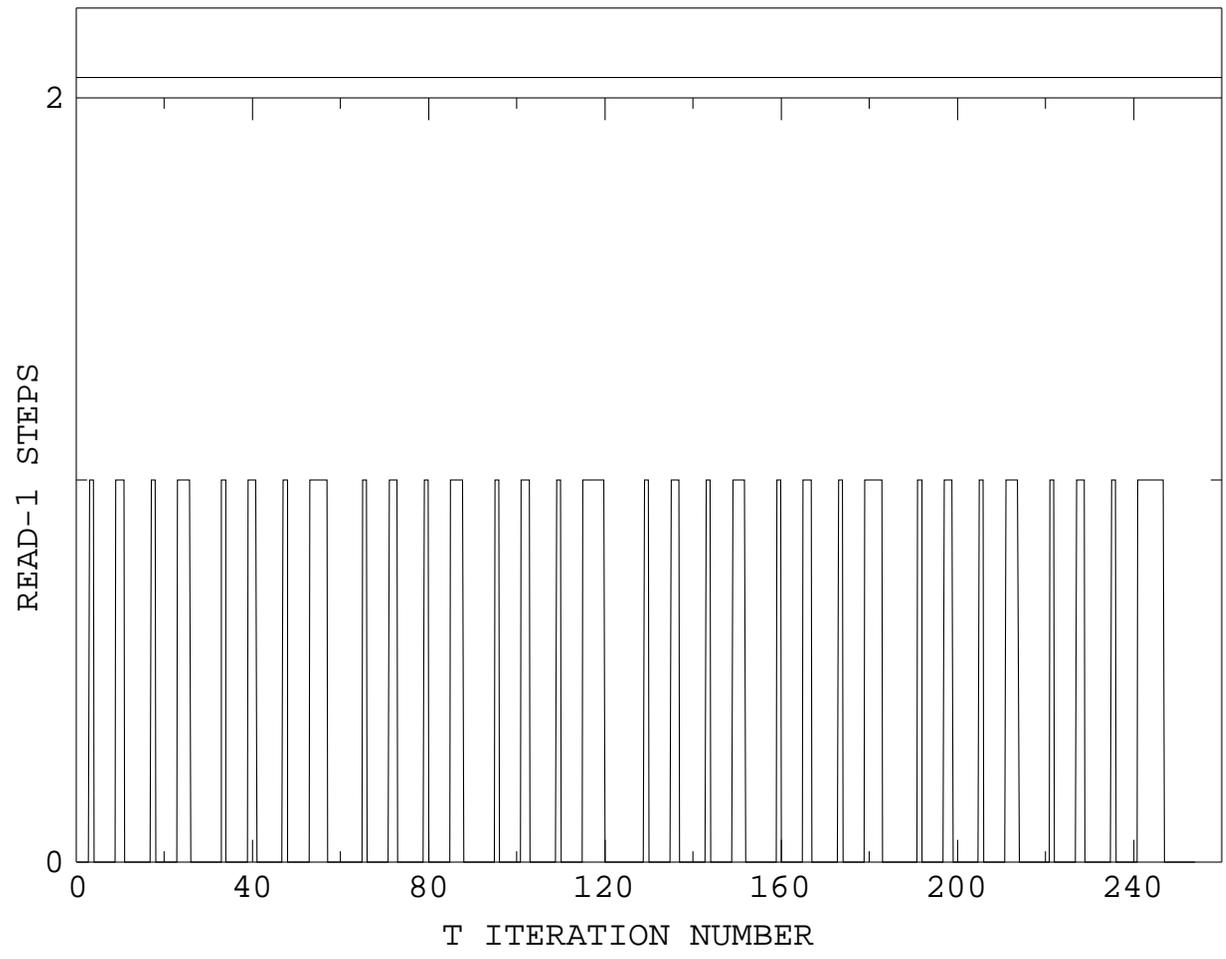
\end{document}